\documentclass[useAMS]{mn2e}
 \usepackage{times}
 \usepackage{epsfig}
 \usepackage{amssymb}

 \title[Three new stable L$_5$ Mars Trojans]
       {Three new stable L$_5$ Mars Trojans}

 \author[C. de la Fuente Marcos and R. de la Fuente Marcos]
        {C.~de~la~Fuente~Marcos\thanks{E-mail: nbplanet@fis.ucm.es}
         and
         R. de la Fuente Marcos \\
         Universidad Complutense de Madrid,
         Ciudad Universitaria, E-28040 Madrid, Spain}
 \date{Accepted 2013 February 25.
       Received 2013 February 4;
       in original form 2012 September 24}
 \pubyear{2013}
 
\begin{document}
  \maketitle

  \begin{abstract}
     Mars was second to Jupiter in being recognized as the host of a population of 
     Trojan minor bodies. Since 1990, five asteroids -- 5261 Eureka, (101429) 1998 
     VF$_{31}$, (121514) 1999 UJ$_7$, 2001 DH$_{47}$ and (311999) 2007 NS$_{2}$ -- 
     have been identified as Mars Trojans, one L$_4$ and four L$_5$. Dynamical and 
     spectroscopic evidence suggests that some Mars Trojans may be remnants of the 
     original planetesimal population that formed in the terrestrial planets region. 
     Here, we revisit the long-term dynamical evolution of the previously known Mars 
     Trojans and show~that 2011 SC$_{191}$, 2011 SL$_{25}$ and 2011 UN$_{63}$ are 
     also trailing (L$_5$) Mars Trojans. They appear to~be as stable as Eureka and 
     may have been Trojans over the age of the Solar system. The fact~that five 
     Trojans move in similar orbits and one of them is a binary may point to the 
     disruption~of~a larger body early in the history of the Solar system. Such a 
     catastrophic event may also~explain the apparently strong asymmetry in terms of 
     number of objects (one versus seven)~between the L$_4$ and L$_5$ regions. Future 
     spectroscopic observations should be able to reject or confirm~a putative common 
     chemical signature that may lend further support to a collisional scenario.
  \end{abstract}


  \section{Introduction}
     Trojans are minor bodies that share the semimajor axis of their host body but they may have different eccentricity and inclination. In 
     a frame of reference rotating with the host, they move in the so-called tadpole orbits around the Lagrangian equilateral points L$_4$ 
     and L$_5$; L$_4$ is located on the orbit of the host object at some 60$^{\circ}$ ahead or east of the host, and L$_5$ is some 
     60$^{\circ}$ west. In other words and for a Trojan object, the relative mean longitude $\lambda_r = \lambda - \lambda_H$ oscillates 
     around the values +60$^{\circ}$ (L$_4$) or -60$^{\circ}$ (or 300$^{\circ}$, L$_5$), where $\lambda$ and $\lambda_H$ are the mean 
     longitudes of the Trojan and the host body, respectively. The first Trojan object, 588 Achilles, was discovered by M. Wolf in 1906 at 
     Jupiter's L$_4$ point (Einarsson 1913). Since 1990, Mars is second to Jupiter in being recognized as the host of a population of Trojan 
     minor bodies (see, for example, Marzari et al. 2002 for a review). The list of Mars Trojans currently includes 5261 Eureka (Bowell et 
     al. 1990; Holt et al. 1990; Mikkola et al. 1994), (101429) 1998 VF$_{31}$ (Ticha et al. 1998; Tabachnik \& Evans 1999), (121514) 1999 
     UJ$_7$ (Rivkin et al. 2003), 2001 DH$_{47}$ (Scholl, Marzari \& Tricarico 2005) and (311999) 2007 NS$_{2}$ (Rodriguez et al. 2007; 
     Izidoro, Winter \& Tsuchida 2010; Schwarz \& Dvorak 2012). Right after the discovery of Eureka, it was suggested that Mars Trojans may 
     have been captured fairly late in Solar system history (Bowell et al. 1990) but soon after, the primeval Mars Trojan scenario started 
     to gain support from both dynamical (long-term integrations) and spectroscopic data (Mikkola et al. 1994; Howell et al. 1995). Today, 
     it is widely accepted that some Mars Trojans may be primordial bodies, perhaps the only surviving examples of the planetesimal 
     population that formed in the terrestrial planets region (e.g., Connors et al. 2005; Scholl et al. 2005). 
     \hfil\par
     Any claim regarding the putative ancient nature of all or part of the current Mars Trojan populations must be supported by dynamical
     stability analyses. The stability of Martian Trojans was first studied by Mikkola \& Innanen (1994). For a simulated time of 100~Myr, 
     they found two regions of stability depending on the orbital inclination, $i$; only orbits with $i\in$ (15, 30)$^{\circ}$ or $i\in$ 
     (32, 44)$^{\circ}$ appeared to be long-term stable. Tabachnik \& Evans (1999, 2000) found that the inclination ranges in Mikkola \& 
     Innanen (1994) ensured stability around the Lagrangian point L$_4$ but that around L$_5$, the range of stable inclinations was $i\in$ 
     (15, 40)$^{\circ}$. In their 100~Myr simulations, both Eureka and 101429 were deeply settled into the L$_5$ stable region; they also 
     suggested a primordial origin for these two objects. Focusing on the role of secular resonances in the dynamical evolution of Mars 
     Trojans, Brasser \& Lehto (2002) found that several secular resonances with the Earth, Mars and Jupiter were able to explain the 
     unstable inclination ranges found by previous studies. Scholl et al. (2005) further explored this topic to conclude that secular 
     resonances make Trojan orbits with $i<15^{\circ}$ as well as those with $i\sim30^{\circ}$ unstable. Orbits with $i>35^{\circ}$ are 
     also unstable due to the Kozai resonance (Kozai 1962). In their 4.5~Gyr calculations, they also found that the dynamical half-life of 
     Mars Trojans in the most stable regions is of the order of the age of the Solar system and that the Yarkovsky effect has negligible 
     role for objects larger than about 10~m. All known Mars Trojans have relatively small sizes with diameters of order of 1~km (Trilling 
     et al. 2007). Numerical modelling (Tabachnik \& Evans 1999, 2000) predicts that the number of Mars Trojans with diameter $>$ 1~km could 
     be as high as 30.
     \hfil\par
     Here, we search for Mars Trojan candidates among known asteroids with relative semimajor axis $|a - a_{\sf Mars}| <$ 0.001 au, then 
     perform $N$-body calculations to confirm or reject their Trojan nature. In this Letter, we show that the minor bodies 2011 SC$_{191}$, 
     2011 SL$_{25}$ and 2011 UN$_{63}$ are stable trailing (L$_5$) Mars Trojans. This Letter is organized as follows: in Section 2, we 
     briefly outline our numerical model. Section 3 focuses on currently known Mars Trojans, reviewing their dynamical status. Section 4 
     presents our results for the objects studied here. These results are discussed and our conclusions summarized in Section 5.
%
%
  \begin{table*}
   \fontsize{8}{10pt}\selectfont
   \tabcolsep 0.35truecm
   \caption{Heliocentric Keplerian orbital elements of known Mars Trojans. Values include the 1-$\sigma$ uncertainty.
            (Epoch = JD2456200.5, 2012-Sep-30.0; J2000.0 ecliptic and equinox. Source: JPL Small-Body Database and AstDyS-2.)
           }
   \resizebox{\linewidth}{0.12\linewidth}{
   \begin{tabular}{ccccccc}
    \hline
                                                        &   & Eureka & 101429 & 121514 & 2001 DH$_{47}$ & 311999 \\
    \hline
     semimajor axis, $a$ (au)                           & = & 1.523 536 2                   & 1.524 181 4                   & 1.524 424 5                   
                                                            & 1.523 798 6                   & 1.523 756 9                  \\
                                                        &   & $\pm$0.000 000 005            & $\pm$0.000 000 02             & $\pm$0.000 000 12 
                                                            & $\pm$0.000 000 2              & $\pm$0.000 000 03            \\
     eccentricity, $e$                                  & = & 0.064 767 3                   & 0.100 311 5                   & 0.039 184 3                   
                                                            & 0.034 659 6                   & 0.053 978 1                  \\
                                                        &   & $\pm$0.000 000 08             & $\pm$0.000 000 13             & $\pm$0.000 000 12
                                                            & $\pm$0.000 000 6              & $\pm$0.000 000 6             \\
     inclination, $i$ ($^{\circ}$)                      & = & 20.283 11                     & 31.298 20                     & 16.749 27                   
                                                            & 24.398 79                     & 18.620 98                    \\
                                                        &   & $\pm$0.000 007                & $\pm$0.000 02                 & $\pm$0.000 012
                                                            & $\pm$0.000 05                 & $\pm$0.000 02                \\  
     longitude of ascending node, $\Omega$ ($^{\circ}$) & = & 245.069 82                    & 221.331 79                    & 347.396 04                 
                                                            & 147.432 50                    & 282.499 55                   \\
                                                        &   & $\pm$0.000 014                & $\pm$0.000 02                 & $\pm$0.000 02
                                                            & $\pm$0.000 04                 & $\pm$0.000 04                \\ 
     argument of perihelion, $\omega$ ($^{\circ}$)      & = & 95.414 85                     & 310.573 87                    & 48.417 49                   
                                                            & 17.615                        & 176.873 4                    \\
                                                        &   & $\pm$0.000 07                 & $\pm$0.000 11                 & $\pm$0.000 2
                                                            & $\pm$0.002                    & $\pm$0.0002                  \\
     mean anomaly, $M$ ($^{\circ}$)                     & = & 236.388 24                    & 32.830 35                     & 321.501 44                  
                                                            & 53.521                        & 120.741 26                   \\
                                                        &   & $\pm$0.000 07                 & $\pm$0.000 13                 & $\pm$0.000 2
                                                            & $\pm$0.003                    & $\pm$0.000 14                \\
     absolute magnitude, $H$                            & = & 16.1                          & 17.4                          & 16.9                     
                                                            & 19.7                          & 17.8                         \\
    \hline
   \end{tabular}
   }
   \label{elements}
  \end{table*}
%
%
%
%
     \begin{figure*}
       \centering
        \includegraphics[width=\linewidth,height=0.36\linewidth]{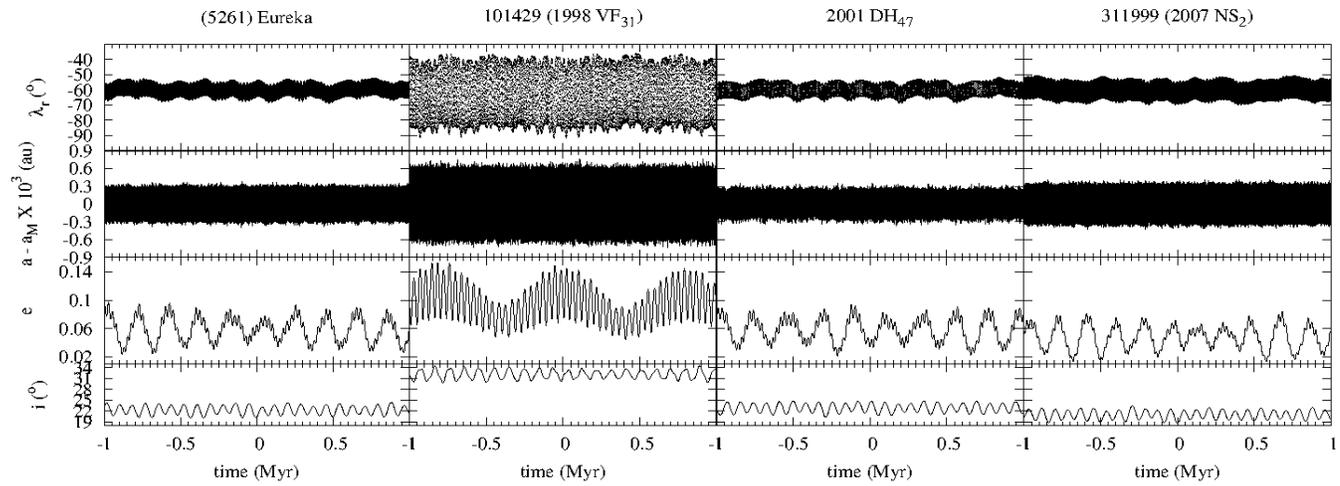}
        \caption{Time evolution of various parameters for all the previously known L$_5$ Mars Trojans. 
                }
        \label{known1}
     \end{figure*}
%
%

  \section{Numerical model}
     In order to confirm the current Trojan nature of our candidates, we perform $N$-body calculations in both directions of time ($\pm$2 
     Myr) with the Hermite integrator (Makino 1991; Aarseth 2003), in a model Solar system which includes the perturbations by the eight 
     major planets and treat the Earth and the Moon as two separate objects; it also includes the barycentre of the dwarf planet 
     Pluto-Charon system and the five largest asteroids, 1 Ceres, 2 Pallas, 4 Vesta, 10 Hygiea and 31 Euphrosyne. Our calculations consider 
     point, constant mass objects orbiting in a conservative system; therefore, relativistic effects are ignored and no modeling of the 
     Yarkovsky and Yarkovsky-O'Keefe-Radzievskii-Paddack (YORP) effects (see, e.g., Bottke et al. 2006) is attempted. The standard version 
     of this direct $N$-body code is publicly available from the IoA web site;\footnote{http://www.ast.cam.ac.uk/$\sim$sverre/web/pages/nbody.htm} 
     additional details can be found in de la Fuente Marcos \& de la Fuente Marcos (2012). In order to validate our results, we have also 
     calculated the evolution of the previously known Mars Trojans (see Figs \ref{known1} and \ref{known2}). Our orbital integrations use 
     initial conditions (positions and velocities in the barycentre of the Solar system referred to the JD2456200.5 epoch) provided by JPL's 
     \textsc{Horizons} system (Giorgini et al. 1996; Standish 1998). In all the figures, $t$ = 0 coincides with the JD2456200.5 epoch. 
     Relative errors in the total energy at the end of the simulations are $< 5 \times 10^{-15}$. In addition to the calculations completed 
     using the nominal orbital elements in Tables \ref{elements} and \ref{elementsnew}, we have performed 20 control simulations (for each 
     object) using sets of orbital elements obtained from the nominal ones within the accepted uncertainties (3$\sigma$). The sources of the 
     Heliocentric Keplerian osculating orbital elements and uncertainties are the JPL Small-Body Database\footnote{http://ssd.jpl.nasa.gov/sbdb.cgi} 
     and the AstDyS-2 portal.\footnote{http://hamilton.dm.unipi.it/astdys/} The results of these control calculations are consistent with 
     those from the nominal ones. In order to study the long term stability of these newly found Trojans we use the Regularized Mixed 
     Variable Symplectic (RMVS) integrator (\textsc{swift-rmvs3}) which is part of the \textsc{SWIFT} package (Levison \& Duncan 1994). The 
     \textsc{SWIFT} package is available from H. Levison web site.\footnote{http://www.boulder.swri.edu/$~$hal/swift.html} The physical 
     model here is similar to the previous one but the Earth and the Moon are replaced by the barycentre of the system. One group of 
     calculations does not include Mercury or the Pluto-Charon system, uses a timestep of 9 d and has been followed for $\pm$4.5 Gyr with 
     six control orbits per object; the second group includes Mercury and the Pluto-Charon system, uses a timestep of 3 d and runs for 
     $\pm$2.25 Gyr. Relative errors in the total energy are now $< 1.5 \times 10^{-8}$. Results from RMVS and Hermite are consistent within 
     the applicable timeframe ($\pm$2 Myr). 

  \section{Known Mars Trojans}
     In this section, we present the results of the short-term integrations of the orbits of the known Mars Trojans (nominal orbital 
     elements in Table \ref{elements}). These calculations are presented here because there are no references including a dynamical analysis 
     of all of them; these results are also used to validate our simulations against previous work. Our results are consistent with those 
     from other authors.
     \hfil\par 
     {\bf 5261 Eureka}. Discovered by D. H. Levy and H. E. Holt on 1990 June 20 during the course of the Mars and Earth-crossing Asteroid 
     and Comet Survey conducted by E. M. and C. S. Shoemaker (Holt et al. 1990), the first known images of this object date back to 1979 
     November and its dynamics has been studied in detail for more than 20 yr. Its diameter is 1.3-2.6 km (Trilling et al. 2007; Burt 2012), 
     its rotational period is 2.69 h and it has a secondary companion with an orbital period of 16.93 h (Burt 2012). The period of its L$_5$ 
     Trojan motion varies between 1300 and 1400 yr (Mikkola et al. 1994) as a result of the changes in inclination; the orbit is stabilized 
     by the perturbation of the neighbouring planets. Its relative mean longitude is modulated with a major periodicity of nearly 200\,000 
     yr, which is the circulation period for the longitude of perihelion, $\Omega + \omega$. Our calculations confirm that it is an L$_5$ 
     Mars Trojan as well as all the results in Mikkola et al. (1994), see Fig. \ref{known1}. Its $\lambda_r$ oscillates around -60$^{\circ}$ 
     with an amplitude of 11$^{\circ}$ and a libration period of 1365 yr. Its eccentricity oscillates mainly due to secular resonances with 
     the Earth and the oscillation in inclination appears to be driven by secular resonances with Jupiter.  
     \hfil\par
     {\bf (101429) 1998 VF$_{31}$}. This 800 m minor body (Trilling et al. 2007) was nominated as L$_5$ Trojan by Tabachnik \& Evans (1999) 
     shortly after its discovery. We also confirm that it is an L$_5$ Mars Trojan. Its $\lambda_r$ oscillates around -60$^{\circ}$ with an 
     amplitude of 45$^{\circ}$ and a libration period of 1425 yr, see Fig. \ref{known1}. The libration amplitude is much larger than that of 
     Eureka and, consistently, its orbit appears to be less stable. As in the previous case, oscillations in eccentricity are driven 
     by the Earth and those in inclination by Jupiter. Fig. \ref{known1} clearly shows that 101429 is not a dynamical analogue of Eureka; 
     the evolution of its $\lambda_r$, $a$ and $e$ is not similar to that of Eureka. The intrinsically different nature of these two objects 
     has been further confirmed by spectroscopic results (Rivkin et al. 2007; Lim et al. 2011). Its origin points to a possible capture 
     event 3.9 Gyr ago (Rivkin et al. 2007). 
     \hfil\par
\begin{figure}
       \centering
        \includegraphics[width=\linewidth,height=0.75\linewidth]{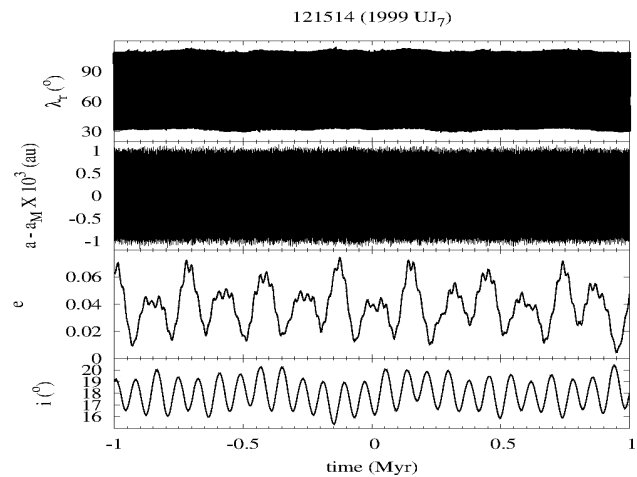}
        \caption{Same as Fig. \ref{known1} but for (121514) 1999 UJ$_{7}$, the only known L$_4$ Trojan.
                }
        \label{known2}
     \end{figure}

     {\bf (121514) 1999 UJ$_7$}. This object is first mentioned as an L$_4$ Mars Trojan by Rivkin et al. (2003); they also point out that
     its visible spectrum is different from those of Eureka or 101429. It is likely a P-asteroid and the largest of the known Mars Trojans
     with a rotational period of 21.2 h (Burt 2012). Our calculations confirm that this object is following a tadpole orbit east of Mars so 
     it is an L$_4$ Mars Trojan. Its $\lambda_r$ oscillates around +60$^{\circ}$ in very wide, asymmetric loops with an amplitude of 
     77$^{\circ}$, the largest of all the objects discussed here, and it has a libration period of 1500 yr, see Fig. \ref{known2}. 
     Simulations (Mikkola et al. 1994) show that, for Mars, stable excursions about L$_4$ and L$_5$ as large as 80$^{\circ}$ occur on 
     time-scales of millions of years which is consistent with our own findings. Oscillations in both eccentricity and inclination are also 
     observed.
     \hfil\par
%
%
%
%
     {\bf 2001 DH$_{47}$}. We also confirm that this object is a robust L$_5$ Mars Trojan. Its $\lambda_r$ oscillates around -60$^{\circ}$ 
     with an amplitude of 11$^{\circ}$ and a libration period of 1365 yr. Together with Eureka, it has the smallest libration amplitude of 
     all the objects studied here. Its orbital behaviour is also very similar to that of Eureka, see Fig. \ref{known1}. 
     \hfil\par  
     {\bf (311999) 2007 NS$_{2}$}. Again, we face a robust L$_5$ Mars Trojan. Its $\lambda_r$ oscillates around -60$^{\circ}$ with an 
     amplitude of 14$^{\circ}$ and a libration period of 1310 yr. Its orbit is similar to that of Eureka and 2001 DH$_{47}$, see Fig. 
     \ref{known1}. Out of all Mars Trojans, it currently has the smallest relative (to Mars) semimajor axis, 5.9$\times10^{-5}$ au. 
%
%
     \begin{table}
      \fontsize{8}{10pt}\selectfont
      \tabcolsep 0.35truecm
      \caption{Heliocentric Keplerian orbital elements of the new objects studied in this research.
               Values include the 1-$\sigma$ uncertainty when available.
               (Epoch = JD2456200.5, 2012-Sep-30.0; J2000.0 ecliptic and equinox.
               Source: JPL Small-Body Database and AstDyS-2.)
              }
      \resizebox{\linewidth}{0.27\linewidth}{
      \begin{tabular}{ccccc}
       \hline
                                &   & 2011 SC$_{191}$  & 2011 SL$_{25}$ & 2011 UN$_{63}$ \\
       \hline
        $a$ (au)                & = & 1.523 809 1      & 1.523 931 0  & 1.523 782       \\
                                &   & $\pm$0.000 000 5 &              & $\pm$0.000 002  \\
        $e$                     & = & 0.044 121 2      & 0.114 465 7  & 0.064 627       \\
                                &   & $\pm$0.000 001 3 &              & $\pm$0.000 007  \\
        $i$ ($^{\circ}$)        & = & 18.744 79        & 21.497 41    & 20.363 2        \\
                                &   & $\pm$0.000 05    &              & $\pm$0.000 3    \\
        $\Omega$ ($^{\circ}$)   & = & 5.799 81         & 9.431 98     & 223.573 0       \\
                                &   & $\pm$0.000 03    &              & $\pm$0.000 2    \\
        $\omega$ ($^{\circ}$)   & = & 196.400          & 53.252 46    & 165.241         \\
                                &   & $\pm$0.005       &              & $\pm$0.005      \\
        $M$ ($^{\circ}$)        & = & 4.422            & 146.943 43   & 192.529         \\
                                &   & $\pm$0.006       &              & $\pm$0.004      \\
        $H$                     & = & 19.3             & 19.5         & 19.9            \\
       \hline
      \end{tabular}
      }
      \label{elementsnew}
     \end{table}
%
%
%
%
     \begin{figure*}
       \centering
        \includegraphics[width=\linewidth,height=0.36\linewidth]{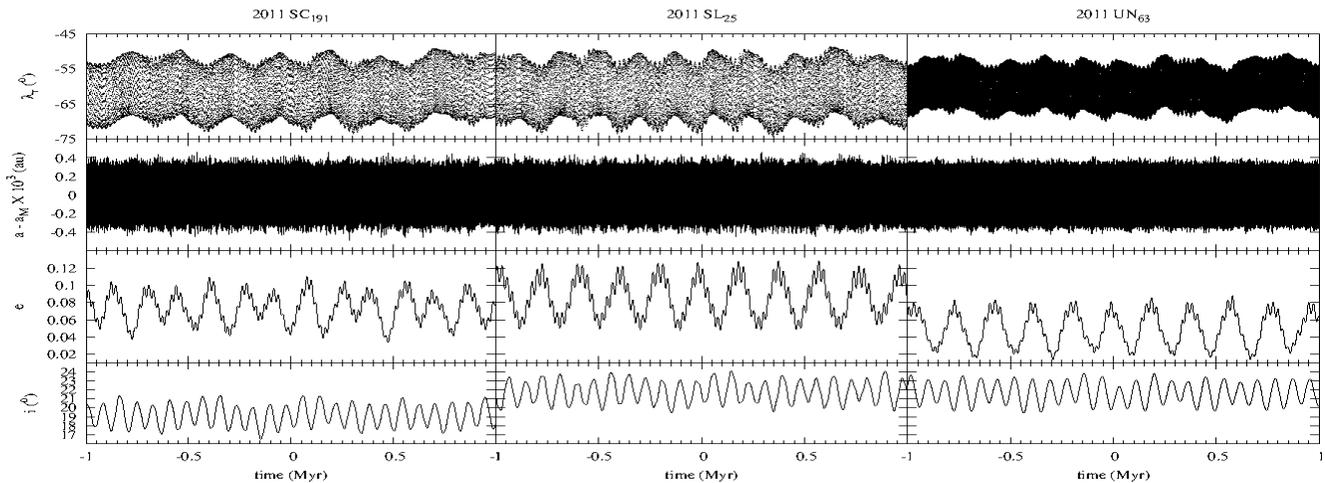}
        \caption{Same as Figs. \ref{known1} and \ref{known2} but for 2011 SC$_{191}$, 2011 SL$_{25}$ and 2011 UN$_{63}$.
                }
        \label{all}
     \end{figure*}
%
%
  \section{New Mars Trojans}
     The three new Trojans follow tadpole orbits around Mars' L$_5$ and, as in previous cases, they exhibit oscillations in both 
     eccentricity and inclination, see Fig. \ref{all}. Their distances to the Earth are always $> 0.4$ au and their arguments of perihelion  
     $\omega$ circulate.
     \hfil\par
\begin{figure}
       \centering
        \includegraphics[width=1.06\linewidth,height=0.6\linewidth]{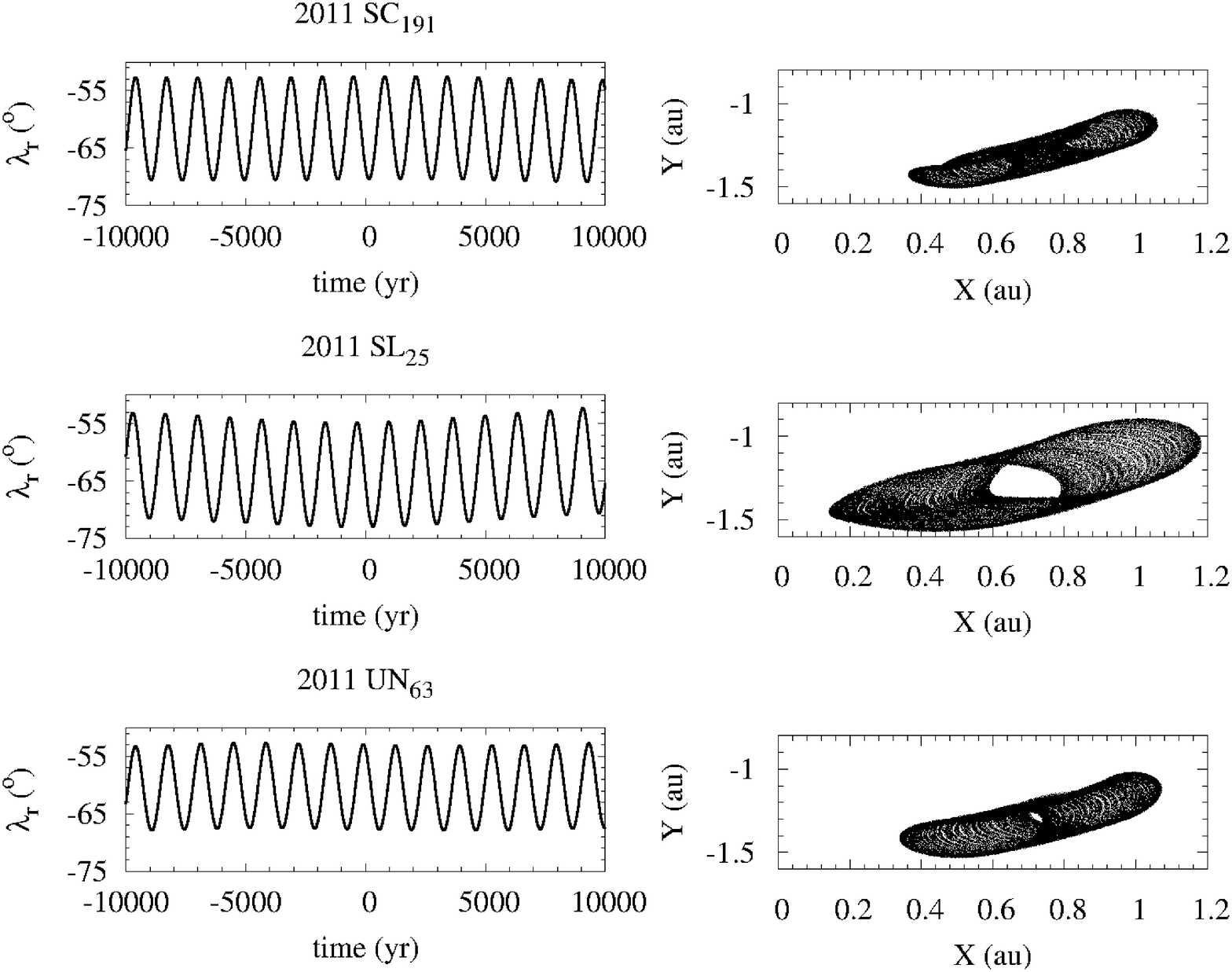}
        \caption{Left-hand panels: details of the evolution of the relative mean longitude for the three objects presented here in the time
                 interval (-10\,000, 10\,000) yr. Right-hand panels: one tadpole loop starting at $t$ = 0 yr in the coordinate system 
                 corotating with Mars; tadpole loops are the superposition of multiple epicyclic loops.
                }
        \label{orbits}
     \end{figure}

     {\bf 2011 SC$_{191}$}. It was originally discovered on 2003 March 21 by the Near-Earth Asteroid Tracking (NEAT) project at Palomar and 
     named 2003 GX$_{20}$, then was lost but re-discovered on 2011 October 31 by the Mt. Lemmon Survey (Pettarin et al. 2011). Its orbit is 
     robust; 45 observations from the years 2003, 2009 and 2011 with an arc length of 3146 d have been used to compute it. Its orbital 
     inclination is significant ($\sim19^{\circ}$) but its eccentricity is low (0.04); both values are similar to those of known Mars 
     Trojans. It is a relatively bright object, suitable for spectroscopy, with a typical apparent visual magnitude of nearly 20. Our 
     calculations indicate that it is a stable L$_5$ Mars Trojan, see Fig. \ref{all}. Its relative mean longitude oscillates around 
     -60$^{\circ}$ with an amplitude of 18$^{\circ}$ and a libration period of 1300 yr, see Fig. \ref{orbits}. These values are comparable 
     to those of the first Mars Trojan, Eureka (see Section 3); the evolution of their orbits is also very similar, see Fig. \ref{known1}. 
     The orbit appears to be very stable on Myr time-scales. As in the case of Eureka, $\lambda_r$ is modulated with a major periodicity of 
     nearly 200\,000 yr. 
     \hfil\par
     {\bf 2011 SL$_{25}$}. It was discovered on 2011 September 21 at the Alianza S4 Observatory on Cerro Burek in Argentina. Its orbit has 
     been computed using 76 observations with an arc length of 42 d; even if somewhat reliable, it is in need of further observations in 
     order to make it as robust as that of 2011 SC$_{191}$. Because of its larger eccentricity, it can get brighter than 2011 SC$_{191}$ at
     nearly 19 mag. Our calculations indicate that it is a stable L$_5$ Mars Trojan, see Fig. \ref{all}. Its $\lambda_r$ oscillates around 
     -60$^{\circ}$ with an amplitude of 18$^{\circ}$ and a libration period of 1400 yr, see Fig. \ref{orbits}. 
     \hfil\par
     {\bf 2011 UN$_{63}$}. It was originally discovered on 2009 September 27 by the Mt. Lemmon Survey and named 2009 SA$_{170}$. After 
     being lost, it was re-discovered on 2011 October 21, again by the Mt. Lemmon Survey. Its orbit is well known; 64 observations with an 
     arc length of 793 days. Its orbital behaviour is similar to that of 2011 SC$_{191}$. Our calculations also indicate that it is a stable 
     L$_5$ Mars Trojan, perhaps the most stable of this set of three, see Fig. \ref{all}. Its $\lambda_r$ oscillates around -60$^{\circ}$ 
     with an amplitude of 14$^{\circ}$ and a libration period of 1350 yr, see Fig. \ref{orbits}, which are comparable to those of Eureka 
     (see Section 3).
%
%
%
%
     
  \section{Discussion and conclusions}
\begin{figure}
       \centering
        \includegraphics[width=\linewidth,height=1.0\linewidth]{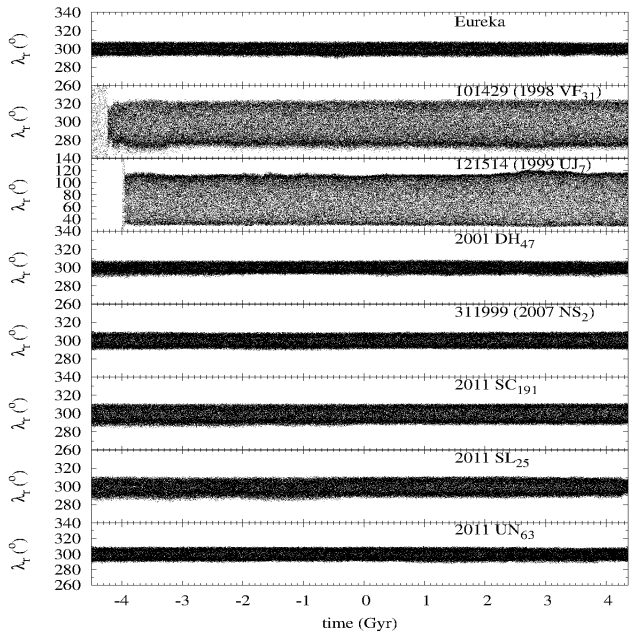}
        \caption{The long-term evolution of the libration amplitude of all known Mars Trojans. All the objects appear to be stable on 
                 Gyr time-scales.
                }
        \label{lt}
     \end{figure}

     The simulated time span studied so far (4 Myr) is not long enough to answer the astrophysically critical question of whether the newly
     found Trojans may have been trapped at the Lagrangian L$_5$ point of Mars since the formation of the Solar system. To investigate the 
     long-term stability of the three objects we use the \textsc{swift-rmvs3} integrator. Our $\pm$4.5 Gyr integrations indicate that the
     lifetime of the Trojan orbits of the new objects is equivalent to or longer than the Solar system age (see Fig. \ref{lt}). They all
     could be primordial objects. We also confirm the long-term stability of the four Mars Trojans studied by Scholl et al. (2005) and 
     conclude that 311999 is also stable on Gyr time-scales. However, our calculations (including control orbits) suggest that 101429 and 
     121514 are not primordial but were captured about 4 Gyr ago. Some of the control orbits associated with 2011 SL$_{25}$ are not stable 
     over the entire simulated time. Our $\pm$2.25 Gyr integrations show stability for all the eight objects.
     \hfil\par
%
%
%
%
     In this Letter, we have presented solid dynamical evidence in the form of $N$-body numerical integrations for three new stable Mars 
     Trojans and also provided a comprehensive summary of the dynamics of the previously known objects. After this study, the list of robust 
     Mars Trojans includes eight objects. Unfortunately and with the exception of the first three objects, there is no consensus among Solar 
     system researchers on the robust Trojan nature of 2001 DH$_{47}$ and 311999 in spite of both having reliable orbits. Even the MPC still 
     lists just three objects as genuine Mars Trojans.\footnote{http://www.minorplanetcenter.net/iau/lists/MarsTrojans.html} The $N$-body 
     calculations completed for this research clearly show that the eight objects are long-term stable, at least for several Gyr. We also 
     conclude that 101429 is by no means a dynamical analogue of Eureka -- their orbital evolution being too different -- which is consistent 
     with the rather different spectroscopic signature found by Rivkin et al. (2007). Two of the three new L$_5$ Trojans move in orbits very 
     similar to that of Eureka which, in the absence of surface composition data, may point to them (and several of the previously known 
     Trojans) being the by-products of the disruption of a larger body. Eureka and 101429 could be collisional fragments of larger bodies as 
     they are highly differentiated (Rivkin et al. 2007). The binarity and relatively fast rotation of Eureka (Burt 2012) as well as the 
     apparently strong asymmetry between the size of the L$_4$ and L$_5$ populations (one versus seven) also point in that direction. 
     Although our calculations indicate that most of the currently known and new Mars Trojans are dynamical analogues of Eureka, 
     spectroscopy should settle the question of them also having a common chemical signature with Eureka, an R chondrite (Lim et al. 2011). 
     If, as suggested, they are fragments of a large body, further objects with similar properties should be found (Marzari et al. 1997). 
     Strategies to search for additional Mars Trojans have been recently discussed by Todd et al. (2012).

  \section*{Acknowledgements}
     We would like to thank S.~Aarseth, H.~Levison and M.~Duncan for providing the codes used in this research and the referee, F.~Marzari, 
     for his constructive and very useful report. This work was partially supported by the Spanish 'Comunidad de Madrid' under grant CAM 
     S2009/ESP-1496 (Din\'amica Estelar y Sistemas Planetarios). We also thank M.~J.~Fern\'andez-Figueroa, M.~Rego Fern\'andez and the 
     Department of Astrophysics of the Universidad Complutense de Madrid (UCM) for providing excellent computing facilities. Most of the 
     calculations and part of the data analysis were completed on the 'Servidor Central de C\'alculo' of the UCM and we thank S.~Cano 
     Als\'ua for his help during this stage. In preparation of this Letter, we made use of the NASA Astrophysics Data System, the ASTRO-PH 
     e-print server and the MPC data server.

  \section*{Note added in press}
     After this work was accepted by MNRAS Letters, a relevant paper was submitted by A. A. Christou to astro-ph (astro-ph/1303.0420). He
     arrives to similar conclusions regarding the three objects discussed here but using different techniques (Christou 2013).

\end{document}